\documentclass[a4paper,dvips,12pt]{article}
\usepackage{epsfig,graphics,graphicx}

\input{tcilatex}

\begin{document}

\title{Light-front time picture of the Bethe-Salpeter equation}
\author{Tobias Frederico$^{a}$, J.H.O.Sales$^{b}$, B.V.Carlson$^{a}$ and P.
U. Sauer$^{c}$. \\
$^{a}$Instituto Tecnol\'{o}gico de Aerona\'{u}tica, CTA, 12228-900\\
S\~{a}o Jos\'{e} dos Campos, Brazil.\\
$^{b}$Instituto de F\'{\i}sica Te\'{o}rica, 01405-900 S\~{a}o Paulo, Brazil.%
\\
$^{c}$Institute for Theoretical Physics, \\
University Hannover, D-30167 Hannover, Germany.}
\maketitle

\begin{abstract}
We show the light-front representation of the field theoretical
Bethe-Salpeter equation (BSE) in the ladder approximation using the quasi
potential reduction. We discuss the equivalence of the covariant ladder
Bethe-Salpeter equation with an infinite set of coupled equations for the
Green's functions of the different light-front Fock-states.
\end{abstract}

\section{Introduction}

Recently, we have proposed a quasi-potential reduction of the
field-theoretical Bethe-Salpeter equation to study two-particle bound states
on the light-front for composite systems of bosons and fermions \cite{j1}.
The reduction scheme eliminates the relative light-front time between the
particles, and the global propagation of the intermediate system appears
leading to Dirac's \cite{dirac} concept of representing the dynamics of the
quantum system for light-front times $x^+=t+z$. We have derived two-body
equations for which the effective interaction is irreducible with respect to
the light-front two-body propagation. The global light-front time projection
makes explicit the intermediate propagation within higher Fock-space
sectors, and the mixing with the valence component of the wave-function. The
effective two-body equation for the valence component of the light-front
wave-function from the quasi-potential reduction has an effective
interaction which contains the coupling with higher-Fock state components.
In lowest order, the two-body bound-state equation is the one first derived
by Weinberg \cite{wei}.

In view of the possible application of the proposed scheme to study the
structure of the bound-state, for example, with electromagnetic probes \cite%
{j1,miller02}, it is timely to reveal the physical aspects of the dynamics
generated by the quasi-potential reduction with respect to the mixing of the
lowest Fock component with higher Fock components of the light-front wave
function. We exemplify our procedure using a model for interacting bosons $%
\phi_1$, $\phi_2$ and $\sigma$ for which the interaction Lagrangian is given
by, 
\begin{eqnarray}
\mathcal{L}_I=g_S\phi _1^{\dagger }\phi _1\sigma +g_S\phi _2^{\dagger }\phi
_2\sigma .  \label{eq1}
\end{eqnarray}

\section{Projection on the Light-Front Hyperplane}

The two-boson transition matrix $T(K)$ in relativistic field theory, for a
given total four-momentum of the system is the solution of 
\begin{eqnarray}
T(K) = V(K)+V(K)G_0(K)T(K) ,  \label{eq2}
\end{eqnarray}
where $V(K)$ is two-body irreducible. Neglecting self-energy parts the
disconnected Green's function, written in \textit{light-front} coordinates,
e.g., $k_i=(k^-_i:=k^0_i-k^3_i\ , \ k^+_i:=k^0_i+k^3_i \ , \ \vec
k_{i\perp}) $, is given by: 
\begin{eqnarray}
& & \left\langle k_1^{\prime -}\right| G_0(K)\left|k^-_1\right\rangle \ = 
\nonumber \\
&-&\frac1{2\pi} \frac{\delta \left(k^{\prime -}_1-k^{-}_1\right)}{\widehat
k^+_1 (K^+-\widehat k^+_1) \left(k_1^--\frac{\widehat{\vec k}%
^2_{1\perp}+m_1^2-io}{\widehat k^+_1}\right) \left(K^--k_1^--\frac{\widehat{%
\vec k}^2_{2\perp}+m_2^2-io}{K^+-\widehat k^+_1} \right)} \ ,  \label{eq3}
\end{eqnarray}
where the "hat" indicates operator form in the "kinematic" momentum ($k^+_1
, \vec k_{1\perp}$), and $\widehat k^-_{1on}=\frac{\widehat{\vec k}%
_{1\perp}^2+m_1^2}{\widehat k^+_1} $ and $\widehat k^-_{2on}= \frac{ (\vec
K_\perp-\widehat{\vec k}_{1\perp})^2+m_2^2}{K^+-\widehat k^+_1} $. The basis
states for functions of the kinematical light-front momentum are defined by $%
\langle x^-_i \vec x_{i\perp}\left| k^+_i\vec k_{i\perp}\right\rangle=
e^{-\imath(\frac12 k^+_ix^-_i-\vec k_{i\perp} . \vec x_{i\perp})}$, which
are eigenfunctions of the momentum operators ($\widehat k^+_i , \widehat{%
\vec k}_{i\perp}$) and the free energy operator $\widehat k^-_{ion} $. The
states $\left|k^+\vec k_\perp \right \rangle$ form a complete basis in the
space of functions of the kinematical variables, e.g., $\int \frac{%
dk^+d^2k_\perp}{2(2\pi)^3} \left. |k^+\vec k_\perp\rangle \langle k^+\vec k
_\perp \right. | c =1$ with normalization $\langle k^{\prime +}\vec
k^\prime_\perp |k^+\vec k_\perp \rangle=2(2\pi)^3\delta(k^{\prime
+}-k^+)\delta(\vec k^{\prime}_\perp-\vec k_\perp)$.

The global light-front time propagator of the free two-boson system between
the hyperplanes $x_1^+=x_2^+=x^+$ and $x_1^{\prime +}=x_2^{\prime
+}=x^{\prime +}$, is the Fourier transform of 
\begin{eqnarray}
g_0(K)=|G_0(K)|:= \int dk^{\prime -}_1 dk^{ -}_1 \left\langle k_1^{\prime
-}\right| G_0(K)\left|k^-_1\right\rangle \ =\frac{ig_0^{(2)}(K)} {\widehat{k}%
^+_{1}(K^+-\widehat{k}^+_{1}) } \ ,  \label{eq4}
\end{eqnarray}
where $K^+\ > \ 0$ is used without any loss of generality. In Eq. (\ref{eq4}%
) the vertical bar $|$ indicates that the dependence on $k^-_1$ is
integrated out. The bar on the left of the Green's function represents
integration on $k^-_1$ in the bra-state, the bar on the right in the
ket-state.

The free two-body light-front Green's function, $g_0^{(2)}(K)$ in Eq.(\ref%
{eq4}) is a particular case of the light-front Green's function for $N$
particles: 
\begin{eqnarray}
g^{(N)}_0(K)= \left[\prod _{j=1}^N \theta (\widehat k_j^{+})\theta
(K^{+}-\widehat k_j^{+})\right]\left( K^{-}-\widehat K_0^{(N)-}+io
\right)^{-1} ,
\end{eqnarray}
where $\widehat K_0^{(N)-}=\sum_{j=1}^N\widehat{k}_{j\ on}^-$ is the free
light-front Hamiltonian. The difference between the free two-body
light-front Green's function and $g_0(K)$, Eq.(\ref{eq4}), is the
phase-space factor for particles 1 and 2.

The covariant two-body propagator is the solution of 
\begin{eqnarray}
G(K) = G_0(K)+G_0(K)V(K)G(K)= G_0(K)+ G_0(K)T(K)G_0(K) .  \label{eq5}
\end{eqnarray}
It gives the individual time propagation of the two-body system. The
interacting propagator between hyperplanes of $x^+$= const., is given by $%
g(K)\equiv|G(K)|$.

Our goal is the decomposition of the kernel of the integral equation of the
global two-body propagator, $g(K)$, with an effective interaction
irreducible with respect to the light-front propagation of the intermediate
two-body system. For that purpose, we use the quasi-potential reduction of
Ref.\cite{wolja}, where the transition matrix $T(K)$ and the Bethe-Salpeter
amplitude $\left| \Psi _B\right\rangle $ of the covariant BSE can be
obtained with the help of an auxiliary Green's function $\widetilde{G}_0(K)$%
, which we choose as 
\begin{eqnarray}
\widetilde G_0(K):= G_0(K)| g_0^{-1}(K) |G_0(K) \ ,  \label{eq6}
\end{eqnarray}
because it allows to make explicit the two-boson propagation on the global
light-front time in the intermediate states\cite{j1}. The transition matrix
satisfies 
\begin{eqnarray}
T(K) & = & W(K)+W(K)\widetilde{G}_0(K)T(K),  \label{eq7} \\
W(K)&=&V(K)+V(K)[G_0(K)-\widetilde{G}_0(K)]W(K) \ .  \label{eq8}
\end{eqnarray}

We introduce the auxiliary three-dimensional transition matrix 
\begin{eqnarray}
t(K)=g_0(K)^{-1}|G_0(K)T(K)G_0(K)|g_0(K)^{-1} \ .  \label{eq9}
\end{eqnarray}
The interaction in the integral equation for $t(K)$, has the desired
property to be irreducible in respect light-front two-body propagation. The
auxiliary transition matrix is the solution of a three-dimensional integral
equation, derived from Eq.(\ref{eq7}): 
\begin{eqnarray}
t(K)&=&w(K)+w(K)g_0(K)t(K),  \label{eq10} \\
w(K)&=&g_0(K)^{-1}|G_0(K)W(K)G_0(K)|g_0(K)^{-1} \ .  \label{eq11}
\end{eqnarray}
From Eqs. (\ref{eq5}) and (\ref{eq10}), one has: 
\begin{eqnarray}
g(K)= g_0(K)+g_0(K)w(K)g(K) \ .  \label{eq12}
\end{eqnarray}
The dynamics of the interacting two-particle system can be fully described
by its propagation between hyperplanes $x^{+}= x^0+x^3=$ const. in
light-front dynamics \cite{dirac}, and covariant the transition matrix can
be expressed in terms of $t(K)$\cite{j1}.

\section{Light-Front Two-Particle Green's Function}

The integral equation satisfied by the light-front Green's function, derived
from Eq.(\ref{eq12}), is 
\begin{eqnarray}
g^{(2)}(K)= g^{(2)}_0(K)+g^{(2)}_0(K) \nu (K) g^{(2)}(K) \ ,  \label{eq13}
\end{eqnarray}
where $g^{(2)}(K)\equiv -i\widehat\Omega g(K)\widehat\Omega$, $\nu (K)=i
\widehat\Omega^{-1} w(K)\widehat\Omega^{-1}$ and the phase space operator is 
$\widehat\Omega:=\sqrt{\widehat k_1^+(K^+-\widehat k^+_1)}$.

The interaction $w(K)$, Eq. (\ref{eq11}), expanded according to Eq. (\ref%
{eq8}) in first order of the driving term $V(K)$, is given by 
\begin{eqnarray}
w^{(2)}(K)=g_0(K)^{-1}|G_0(K)V(K)G_0(K)|g_0(K)^{-1}\ ,  \label{eq14}
\end{eqnarray}
The matrix element $\langle k^{\prime +}_1 \vec k^{\prime}_{1\perp} |
w^{(2)}(K) | k^+_1 \vec k_{1\perp} \rangle $ is obtained from Eq.(\ref{eq14}%
): 
\begin{eqnarray}
\langle k^{\prime +}_1 \vec k^{\prime}_{1\perp} | w^{(2)}(K) | k^+_1 \vec
k_{1\perp} \rangle &=& \left( ig_S\right) ^2 \frac{\theta (k^{+}_1-k^{\prime
+}_1)}{\left( k^{+}_1-k^{\prime +}_1 \right) }\frac i {K^--K^{(3)-}_0+io} 
\nonumber \\
&+& \left[ k^\prime_1 \leftrightarrow k_1 \right] \ ,  \label{eq15}
\end{eqnarray}
where $K^{(3)-}_0= \frac{\vec k^{\prime 2}_{1\perp}+m_1^2}{k^{\prime +}_1} -%
\frac{(\vec K_\perp-\vec k_{1\perp})^2+m_2^2}{K^+-k^{+}_1} -\frac{ (\vec
k^{\prime}_{1\perp} -\vec k_{1\perp} )^2 +\mu^2}{k^+_1-k^{\prime +}_1}$.

The second order term in the expansion of $w(K)$, is given by 
\begin{eqnarray}
w^{(4)}(K)&=&g_0(K)^{-1}|G_0(K)V(K) G_0(K)V(K)G_0(K)|g_0(K)^{-1}  \nonumber
\\
&-&g_0(K)^{-1}|G_0(K)V(K)\widetilde G_0(K)V(K)G_0(K)|g_0(K)^{-1} \ .
\label{eq16}
\end{eqnarray}
The second term in the r.h.s of Eq.(\ref{eq16}) comes from the iteration of $%
w^{(2)}(K)$, which is $w^{(2)}g_0(K)w^{(2)}$. The subtraction of the
iterated term in Eq.(\ref{eq16}) cancels the corresponding terms, such that
the matrix element $\langle k^{\prime +}_1 \vec k^{\prime}_{1\perp} |
w^{(4)}(K) | k^+_1 \vec k_{1\perp} \rangle $ is two-body irreducible. The
final form is: 
\begin{eqnarray}
& &\langle k^{\prime +}_1 \vec k^{\prime}_{1\perp} | w^{(4)}(K) | k^+_1 \vec
k_{1\perp} \rangle =\frac{(ig_S)^4}{2(2\pi)^3}\int dp^+_1d^2p_{1\perp} \frac{%
\theta(p^+_1)}{p^+_1} \frac{\theta(K^{+}-p^+_1)}{(K^{+}-p^+_1)}  \nonumber \\
&\times & \frac{\theta(k^{\prime +}_1-p^+_1)}{(k^{\prime +}_1-p^+_1)}\frac
i{K^{-}-\frac{\vec p_{1\perp }^2+m_1^2}{p_1^{+}}- \frac{\left(\vec
K_\perp-\vec k^{\prime}_{1\perp}\right)^2+m_2^2} {K^+-k_1^{\prime +}}-\frac{%
\left(\vec k^{\prime}_{1\perp} -\vec p_{1\perp}\right)^2+\mu^2}{k^{\prime
+}_1-p_1^{+}}+io}  \nonumber \\
&\times & \frac i{K^{-}-\frac{\vec k_{1\perp }^2+m_1^2}{k_1^{+}} -\frac{%
\left(\vec K_\perp-\vec k^{\prime}_{1\perp}\right)^2+m_2^2} { K^+-k^{\prime
+}_1} -\frac{\left(\vec k^{\prime}_{1\perp}-\vec p_{1\perp}\right)^2+\mu ^2%
} { k^{\prime +}_1-p_1^{+}}-\frac{\left(\vec p_{1\perp}-\vec
k_{1\perp}\right)^2 +\mu^2}{ p_1^+-k_1^{+}}+io}  \nonumber \\
&\times & \frac{\theta(p^+_1-k^+_1)}{(p^+_1-k^+_1)} \frac i{K^{-}-\frac{\vec
k_{1\perp }^2+m_1^2}{k^{+}_1}-\frac{\left(\vec K_\perp-\vec
p_{1\perp}\right)^2+m_2^2}{K^+-p^{+}_1}-\frac{\left(\vec p_{1\perp}-\vec
k_{1\perp}\right)^2+\mu^2}{p^+_1-k_1^{+}}+io}  \nonumber \\
&+&\left[ k^\prime_1 \leftrightarrow k_1 \right] \ .  \label{eq17}
\end{eqnarray}

Introducing the interaction between light-front states which creates or
destroys a quantum of the intermediate boson, $\sigma$ defined by the matrix
elements 
\begin{eqnarray}
\langle q k_\sigma |v|k\rangle&=& -2(2\pi)^3\delta (q+k_\sigma-k) \frac{g_S}{%
\sqrt{q^+k^+_\sigma k^+}} \theta (k^+_\sigma)  \nonumber \\
\langle q|v| k_\sigma k\rangle &=& -2(2\pi)^3\delta (k+k_\sigma-q) \frac{g_S%
}{\sqrt{q^+k^+_\sigma k^+}} \theta (k^+_\sigma) \ ,  \label{eq18}
\end{eqnarray}
for the model defined by the Lagrangian of Eq.(\ref{eq1}), we can rewrite
the first and second order terms $w^{(2)}$ and $w^{(4)}$, in the form of the
effective interaction terms $\nu^{(2)}(K)$ and $\nu^{(4)}(K)$, respectively,
as: 
\begin{eqnarray}
\langle k^{\prime +}_1 \vec k^{\prime}_{1\perp} | \nu^{(2)}(K) | k^+_1 \vec
k_{1\perp} \rangle = \langle k^{\prime +}_1 \vec k^{\prime}_{1\perp} | v
g_0^{(3)}(K) v| k^+_1\vec k_{1\perp} \rangle \ ,  \label{eq19} \\
\langle k^{\prime +}_1 \vec k^{\prime}_{1\perp} | \nu^{(4)}(K) | k^+_1 \vec
k_{1\perp} \rangle= \langle k^{\prime +}_1 \vec k^{\prime}_{1\perp} |
vg_0^{(3)}(K)vg_0^{(4)}(K)vg_0^{(3)}(K)v | k^+_1 \vec k_{1\perp} \rangle .
\label{eq20}
\end{eqnarray}

Therefore, the effective interaction $\nu (K)$ up to second order in the
expansion is the sum of Eqs. (\ref{eq19}) and (\ref{eq20}): 
\begin{eqnarray}
\nu (K)\approx v\left(g_0^{(3)}(K)
+g_0^{(3)}(K)vg_0^{(4)}(K)vg_0^{(3)}(K)\right)v \ .  \label{eq21}
\end{eqnarray}
We identify in Eq.(\ref{eq21}), the propagation of the intermediate
three-particle system, with corrections up to second order in the coupling
constant. It is easy to imagine that the interacting three-particle Green's
function should be obtained if the expansion of the auxiliary interaction $%
w(K)$ is performed to all orders in the ladder using Eqs.(\ref{eq8}) and (%
\ref{eq11}). Thus, the effective interaction should be $\nu (K)=
vg^{(3)}(K)v $. From Eq.(\ref{eq21}), one sees that $g^{(3)}$ is as function
of the four-particle Green's function, which should be coupled to the
five-body one, and so on. Therefore, one can construct a hierarchy of
coupled equations for Green's functions with any number of particles.

\section{Hierarchy Equations and Summary}

The light-front Green's function of the two-body system expressing the
covariant BS equation is written as: 
\begin{eqnarray}
g^{(2)}(K)&=&g^{(2)}_0(K) +g^{(2)}_0(K)v g^{(3)}(K)vg^{(2)}(K) \ ,  \nonumber
\\
g^{(3)}(K)&=&g^{(3)}_0(K) +g^{(3)}_0(K)v g^{(4)}(K)vg^{(3)}(K) \ ,  \nonumber
\\
g^{(4)}(K)&=&g^{(4)}_0(K) +g^{(4)}_0(K)v g^{(5)}(K)vg^{(4)}(K) \ ,  \nonumber
\\
&& \ . \ . \ .  \nonumber \\
g^{(N)}(K)&=&g^{(N)}_0(K) +g^{(N)}_0(K)v g^{(N+1)}(K)vg^{(N)}(K) \ , 
\nonumber \\
&& \ . \ . \ . \ ,  \label{eq22}
\end{eqnarray}
This set of equations contains all two-body irreducible diagrams with
exception of those including closed loops of bosons $\Phi_1$ and $\Phi_2$
and part of the cross-ladder diagrams. The truncation of the light-front
Fock space allows only two bosons states with two particles $\Phi_1$ and $%
\Phi_2$ in the intermediate states, without any restriction on the number of
bosons $\sigma $, which excludes the complete representation of the crossed
ladder diagrams. It also resembles the iterated resolvent method of Ref.\cite%
{brodsky}. To get the two-body propagator for light-front times in the
covariant ladder approximation, the kernel of the hierarchy equations should
be restricted.

A systematic expansion can be obtained by truncating the light-front Fock
space up to $N$ particles in the intermediate states (boson 1, boson 2 and $%
N-2$ $\sigma$'s) in the set of Eqs.(\ref{eq22}), which amounts to
substituting 
$g^{(N)}(K)\rightarrow g^{(N)}_0(K) \ .$ 
By restricting to up to four-particles in the intermediate state
propagation, we get the following nonperturbative equation for the Green's
function: 
\begin{eqnarray}
g^{(2)}(K)&=&g^{(2)}_0(K) +g^{(2)}_0(K)v g^{(3)}(K)vg^{(2)}(K) \ ,
\label{eq24} \\
g^{(3)}(K)&=&g^{(3)}_0(K) +g^{(3)}_0(K)v g^{(4)}_0(K)vg^{(3)}(K) \ .
\label{eq25}
\end{eqnarray}
The kernel of Eq.(\ref{eq24}) still contains an infinite sum of light-front
diagrams, that are obtained by solving Eq.(\ref{eq25}). To get the ladder
approximation up to order $g_S^4$, or up to second order in the
quasi-potential expansion, with the effective interaction of Eq.(\ref{eq22}%
), only the free and first order terms are kept in Eq.(\ref{eq25}), while
restricting it only to the ladder.

In summary, we discussed the general framework for constructing the
light-front two-body Green's function with quasi-potential reduction of the
Bethe-Salpeter equation in the light-front. We displayed a set of coupled
hierarchy equations which gives the two-body propagator in several cases,
including the ladder approximation. Finally we pointed out that the
truncation in the hierarchy can be performed consistently by approximating
the N-body Green's function with the free operator, i.e., the light-front
N-body intermediate state is accounted in lowest order in the two-body
propagation.

\textit{Acknowledgement}\textbf{: }We would like to thank Funda\c{c}\~{a}o
de Amparo \`{a} Pesquisa do Estado de S\~{a}o Paulo (FAPESP) and the
Conselho Nacional de Desenvolvimento Cient\'{\i}fico e Tecnol\'{o}gico
(CNPq) of Brazil for partial support.

\end{document}